# The Application of Digital Technology and the Learning Characteristics of Generation Z in Higher Education

## Research-in-progress

**Ali Alruthaya**
School of Accounting, Information Systems, and Supply Chain
RMIT University
Victoria, Australia
Email: alis-1234@hotmail.com

**Thanh-Thuy Nguyen**
School of Accounting, Information Systems, and Supply Chain
RMIT University
Victoria, Australia
Email: thuy.nguyen21@rmit.edu.au

**Sachithra Lokuge**
School of Business
University of Southern Queensland
Queensland, Australia
Email: ksplokuge@gmail.com

# Abstract

The Generation Z (Gen Z), or the digital natives have never experienced a life without the internet. In addition, the advancement of digital technologies such as social media, smart mobile technologies, cloud computing, and the Internet-of-things has transformed how individuals perform their day-to-day activities. Especially for Gen Z, the use of digital technology has become an essential part of their daily routine, as a result, challenging the norm. As such, Gen Z displays unique learning characteristics which are different from previous generations. This change opens new avenues for exploring the impact of digital technology on the learning characteristics of Gen Z and possible applications to the higher education environment. By conducting a literature review of 80 studies, this paper presents a comprehensive framework for understanding the influence of digital technologies on the learning characteristics of Gen Z in higher education.

# 1 Introduction

The advancement of digital technologies such as social media, smart mobile technologies, cloud computing, analytics and Internet-of-things has become an essential part of day-to-day life due to their pervasiveness, cost efficiency, ease-of-use and ease-of-configuration (Lokuge and Sedera.2014a; Lokuge and Sedera.2014b; Nylén.2015). Young adults, especially Generation Z (Gen Z), who were born after 1995 (Bell.2013), grew up with these technologies and as a result are familiar with these technologies. Especially in economically developed countries, many young people have acquired excessive technological skills before entering university (Lai and Hong.2015). However, smart mobile and internet have become essential utilities even in developing countries (Silver et al.2019), which means Gen Z is well equipped with technological skills. As a result, higher education sector needs to pay attention to provide better services for Gen Z students. There are several challenges associated with the application of current technology in education, and there is a fair amount of research devoted to understanding the possible downsides of this trend (Lacka et al.2021). However, technologies can enhance many elements of learning and self-learning (Lokuge et al.2020; Rosemann et al.2000; Sedera and Lokuge.2019a). Some researchers believe that, as a result of being immersed in technology, there is a fundamental difference in how Gen Z using technology in learning compared to prior generations (Prensky.2001a).

The Gen Z, also known as 'digital natives' have unique learning styles such as multitasking in learning, and a unique way of thinking and accessing information (Prensky.2001a; Prensky.2001b). As a result, higher education teaching should adapt to these variances in order to suit "more technology-driven, spontaneous, and multisensory" learning methods (Lai and Hong.2015). Therefore, if these statements are shown to be true, digital technologies might have a big impact on curriculum design, teaching in higher education. Despite the above circumstances, studies analysing the connection between digital technology use and the learning characteristics of the Gen Z are scarce. As such, the objective of this paper is to understand the impact of digital technologies in higher education for Gen Z.

The structure of the paper is as follows. Next section provides the details of the method followed in the study. Then, the overview of the sample is provided and proposes the propositions and the research framework based on the extant literature review. Finally, the findings of the analysis are discussed highlighting the contributions to academia, practices, and future research areas.

# 2 Research Method

For the literature review, researchers searched for studies published between the 1st of January 2010 and the 31st of July 2021. To identify the relevant literature, a range of databases were searched including ProQuest Central, ScienceDirect, Emerald, Google Scholar, and University online Library. The keywords used for searching the literature consist of "characteristics of learning/or learning characteristics," "learner characteristics," "digital technology/digital technologies use+ higher education," "Gen Z/Generation Z + higher education/or university(ies)," "Gen Z/Generation Z + characteristics," "Generation Z and iGeneration". Initially, the researchers focused on articles discussing the learning characteristics of Gen Z in general and in higher education, following by the impacts of digital technology use in higher education and their association with Gen Z's learning characteristics.

The initial search resulted in identifying papers from ProQuest Central (772 results), ScienceDirect (42 results), Emerald (344 results), Google Scholar (1,640 results), and University online Library (126 results). Totally there were more than 2,900 results. Next, the researchers assessed all the titles of the papers to remove any unrelated papers. This resulted in including papers published in high-ranked journals including The British Journal of Educational Technology, Computers & Education, Computers in Human Behavior, and Studies in Higher Education journals. In total, the researchers collected 80 papers discussing the learning characteristics of Gen Z and the impact of digital technology use on the sphere of higher education. Finally, through the analysis, the authors derived a framework and the propositions.

# 3 Overview of the Literature Sample

The theoretical foundation of the generation is derived from two major sociologists Karl Mannheim, and Norman Ryder (Ortiz-Pimentel et al.2020). Because the birth year is the most defining elements of a person's identity (Mannheim.2013), each generation has its characteristics and features that differ from the one before it (Eckleberry-Hunt and Tucciarone.2011; Sakdiyakorn et al.2021). In many social sciences and humanities studies, the theory of the generation which classified the Generations to X, Y, and Z, have received a significant attention (Pikhart and Klímová.2020).

## 3.1 Generation Z

Gen Z classification is more related to technology innovations that appeared especially the Internet innovation. There is a difference in determining the birth of Generation Z Such as 1992 (Bell.2013), 1994 to 2010 (Król and Zdonek.2020), born after 1995 and the first digital generation (Shams et al.2020; Smaliukiene et al.2020). According to Poláková and Klímová (2019) generation Z is the first generation to have never experienced life before the Internet. Unsurprisingly, the technology revolution was highlighted as a key effect on this group (Sakdiyakorn et al.2021). Their lives are influenced by its usage, and they cannot imagine living without it since it has become a natural part of their daily existence (Farrell and Phungsoonthorn.2020). Gen Z is thought to be avid consumers of technology and truly digital indigenous people (Goh and Lee.2018). Generation Z is growing up in a world saturated with technology and the Internet, complete with cell phones, video games, and displays (Haddouche and Salomone.2018). Since early childhood, they have been exposed to digital technology and virtual space, and they have become avid Internet users (Puchkova et al.2017). According to Haddouche and Salomone (2018) sending of e-mails, SMS, and likes is an essential component of Gen Z's everyday existence. They can begin a video game with a neighbour and continue it with a person on the opposite side of the world (Haddouche and Salomone.2018). This is why Generation Z is often referred to as Generation C (connected), with the "C" standing for their extensive usage of the Internet and social media for both personal and professional communication (Hardey.2011). They are always linked to networks and are quick in all of their actions, including decision-making (Hernandez-de-Menendez et al.2020a).

It is worth to mention that Generation Z is also known as the digital natives generation, a phrase coined by Prensky (2001a), which refers to the generation born in the digital age (Persada et al.2019). The N generation (Net), the D generation (Digital), the V generation (Viral), and the Google generation are some of the nicknames for Generation Z (Cruz and Díaz.2016). Cruz and Díaz (2016) claim that these nicknames all contain the same denominator, information, and communication technology (ICT). Because of their technological addiction, they are also known as the iGeneration, Facebook Generation, Gen Tech, Switchers, Online Generation and "always clicking" generation (Dolot.2018). In addition, Gen Z has the ability to multitask an incredible amount of tasks at the same time (Demir and Sönmez.2021; Puchkova et al.2017). Last but not least, this generation is obsessed with finding out what is going on in its own social and family circle (Haddouche and Salomone.2018).

### 3.2 Learning Characteristics of Gen Z

Gen Z students have grown up in an age of technological advancement. As a result, they spend their whole lives surrounded by a range of digital instruments, which have become vital in their everyday lives (Poláková and Klímová.2019). Gen Z students rely on Google, social media, and YouTube as their major resources for learning and performing research (Ashour.2020). They also get knowledge and solutions to their queries from any source on the Internet, such as Wikipedia, YouTube videos, etc. (Ashour.2020). So far, this generation of students have been able to access digital technologies more than other generations (Sakdiyakorn et al.2021).

Because Gen Z students have spent their lives immersed in technology, some academics believe they have a distinctive learning style (Hernandez-de-Menendez et al.2020b). Even though some prior studies exposed that there are no significant differences between the Generation Z and previous generations in learning characteristics (Lai and Hong.2015; Thinyane.2010; Thompson.2013), recent researches confirmed that the differences exist (Hernandez-de-Menendez et al.2020b). For example, one of the distinguishing features of Generation Z learners is their interest in, and attraction to, the visual nature rather than reading texts without multimedia (Ashour.2020).

Students of Gen Z are observers and they like to observe others perform things before implementing what they have learned and apply (Seemiller and Grace.2017). They also like hands-on learning activities that allow them to instantly apply what they learn in the classroom to real-world situations. Further, they want to know that the principles they are learning can be applied to more than simply a practice scenario (Seemiller and Grace.2017). A recent survey found that more than half of Gen Z students spend 1–3 hours each day on social media for personal reasons (Vizcaya-Moreno and Pérez-Cañaveras.2020).

Social media has thus impacted on the learning characteristics of Gen Z with a need for obtaining fast feedback. As a result of these unique learning features of the Gen Z, researchers recommend adapting teaching methods to be more appropriate for the current generation of learners. Therefore, higher institutions must change their teaching approaches associating with digital technology use to become more visual, interactive, with immediate access to materials, and, most critically, to incorporate technology use and social media/networking into the classroom (Cilliers.2017). As a result, the researchers recommend integrating creative lectures (e.g., the uses of visual graphics or videos) with digital simulation and case studies to increase engagement and learning levels (Vizcaya-Moreno and Pérez-Cañaveras.2020).

These conclusions lead us to our first proposition:

P1: *Gen Z's learning characteristics necessitate advanced digital technology use in higher education learning.*

## 3.3 Digital Technology in Higher Education

Digital learning in higher education (HE) include interactive learning resources, digital content learning, learning and teaching software or simulations that involve students in academic material, access to academic publications, online and computer-based evaluation, instructive films, educative articles, and a variety of other tools (Bower.2019). Wikis, blogs, social media, mobile applications, virtual worlds, learning management systems, and the uses of other digital technologies are becoming increasingly common in formal learning environments, particularly in online learning environments such as massive open online courses (MOOCs), and in many of these cases, digital technology is the means by which participants interact (Bower.2019; Lokuge and Sedera.2018; Sedera and Lokuge.2017). Thus, digital technology use in learning in HE includes digital learning platforms, mobile devices, social media, Augmented Reality, etc. M-learning or mobile learning has become more and more popular, because of the low cost of communications and the excellent quality of mobile devices (Park et al.2012), and is a subset of digital learning (Persada et al.2019). Mobile technologies have the ability to break down geographical barriers and convert it into a faceless, virtual world (Sedera and Lokuge.2019b; Sedera and Lokuge.2020) in the classroom (Nguyen et al.2015). Social platforms like Facebook, Twitter and WeChat allow users to openly express their thoughts regarding items in a timely way (Palekar et al.2015; Sedera et al.2016; Shang et al.2017).

Besides, higher education institutions are rapidly embracing digital technologies, such as using the Virtual Learning Environment and Social Media for the students' benefit (Tess.2013), to drive towards "student-centred" strategy (Evans.2014; Tess.2013). The uses of blended learning methods have been encouraged, in which the applications of digital technology can be integrated with traditional education in several levels (Poláková and Klímová.2019; Wang and Tahir.2020). For instance, from a low level such as integrating a digital platform for learning and face-to-face instructors (Poláková and Klímová.2019; (Hernandez-de-Menendez et al.2020a), to a high level such as the implementation of a learning platform in courses, named Kahoot, which combines student response systems (SRS), existing technological infrastructure, with students' own digital devices, their social networking, and games (Wang and Tahir.2020). As such, we derive our second proposition as follows.

*P2: There are several typologies of digital technology use in HE*

| **Level** | **Digital Technology uses** | **References** |
|---|---|---|
| 1 - Low | Digital Platform for learning and face-to-face instructors | Poláková and Klímová (2019), (Hernandez-de-Menendez et al. (2020a), etc. |
| 2 - Medium | Digital platform integrated with social media and mobile devices, and face-to-face instructors | Persada et al. (2019), Bower (2019, Park et al. (2012), Shang et al. (2017), Evans (2014), etc. |
| 3 - High | Digital platform, and social media associated with mobile devices used in Virtual Learning Environment | Nguyen et al. (2015), Shang et al. (2017), Bower (2019), Evans (2014), Tess (2013), Wang and Tahir (2020), etc. |

*Table 1: Digital technology use of the students in higher education*

Obviously there are number of researchers encouraged replacing traditional educational techniques with digital learning (Szymkowiak et al.2021). To enable multimedia-based teaching, digital technology must be strongly integrated into the curriculum, and all learners must have on-demand access to learning resources (Lai and Hong.2015). In addition, Lai and Hong (2015) indicate that the digital technology use in higher education must be considered following the students' learning demographics, such as students' location, students' age, students' gender, students' years, and etc. As such, our third proposition is as follows:

*P3: The digital technology use and Gen Z's learning demographics are positively correlated.*

# 4 Discussion

Based on our findings from the literature review and table 1, the following propositions and a preliminary framework are developed.

P1: Gen Z's learning characteristics necessitate advanced digital technology use in higher education learning.

P2: There are several typologies of digital technology use in HE

P3: The digital technology use and Gen Z's learning demographics are positively correlated.

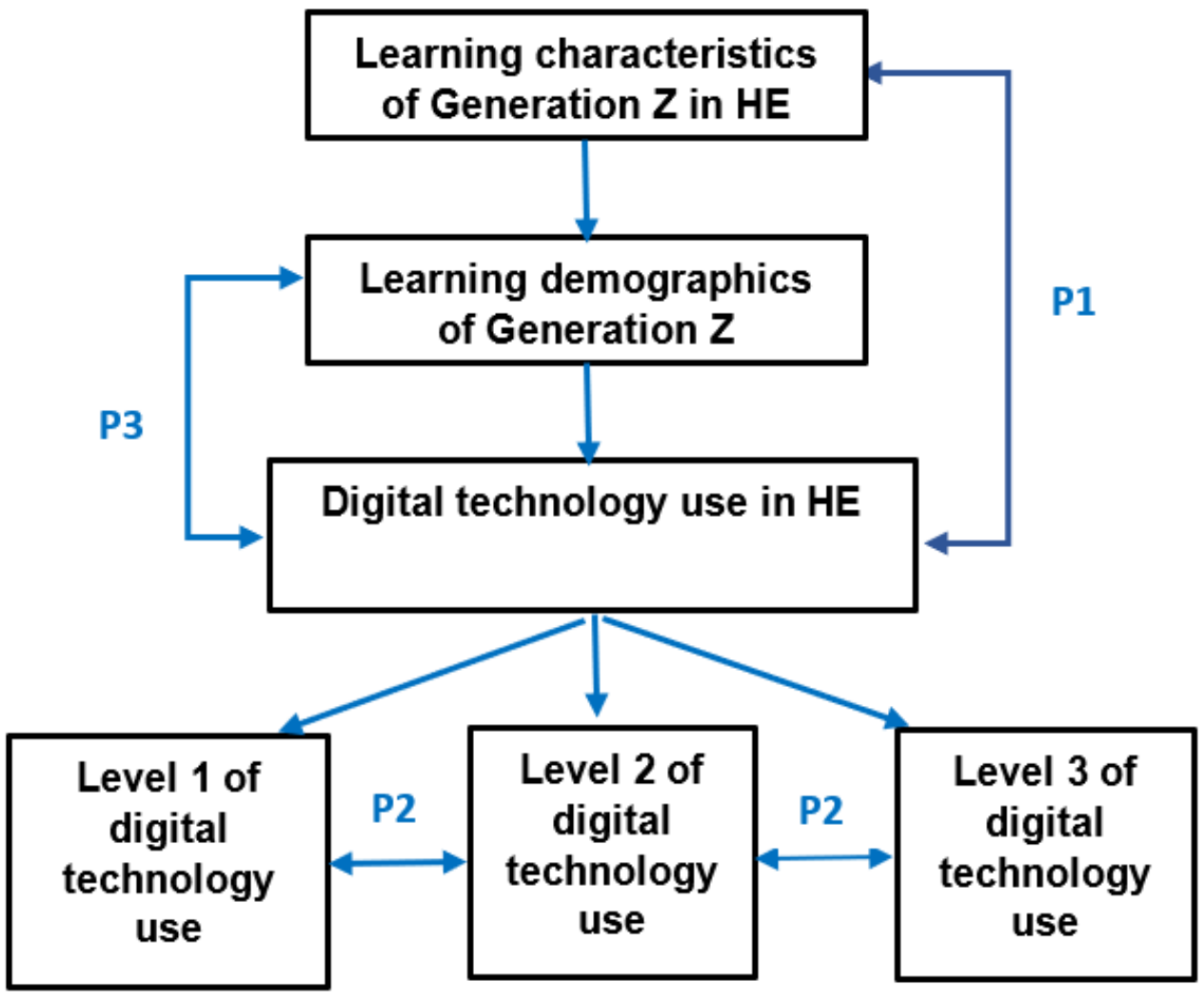

*Figure 1: Theoretical Research Framework*

# 5 Conclusion

This study aims to explore the impact of digital technology on Gen Z in higher education by reviewing previous literature from several disciplines to clarify the future vision of the research. It has been suggested that the learning characteristics of Gen Z in higher education are unique and different from previous generations. Moreover, understanding the connection between the digital technologies use and Gen Z's characteristics will support the policy-making and decision-making process in higher education, both in terms of the right digital technology use and the adoption of blended learning methods.

Obviously, not all regions of the world have the same challenges; some nations might have contextual limits as a result of their educational systems and resources. As such, the framework that we have proposed in this paper will provide guidance for understanding (i) how Gen Z's learning characteristics necessitate advanced digital technology use in HE learning, (ii) typologies of digital technology in HE, and (iii) how the levels of digital technological use are characterised by the learning demographics of Gen Z students. To provide further clarity, future research will be required focusing on conducting empirical research on this topic.

## Copyright